\documentclass{./styles/svproc}
\usepackage{url}
\usepackage{comment}
\usepackage{algorithm}
\usepackage{algpseudocode}
\usepackage{comment}
\usepackage{listings}
\usepackage{cite}
\usepackage{amsmath,amssymb,amsfonts}
\usepackage{graphicx}
\usepackage{textcomp}
\usepackage{xcolor}
\usepackage{pdfpages}

\begin{document}
\mainmatter              
\title{Driver Safety Reward with Cooperative Platooning using Blockchain}
\titlerunning{Driver Safety Reward with Cooperative Platooning using Blockchain}  
%
\author{Sruthi Rachamalla\inst{1} \and Henry Hexmoor\inst{2}}
%
\authorrunning{Sruthi Rachamalla and Henry Hexmoor} 
%
%
\institute{Southern Illinois University, Carbondale, IL 62901, USA,\\
\email{sruthirachamalla@siu.edu and henry.hexmoor@cs.siu.edu }}

\maketitle              

\begin{abstract}
Cooperative driving (or Platooning) focuses on improving the safety and efficiency by connecting two or more vehicles on a road by vehicular communication protocols. The leader is crucial as it manages the platoon, establishes communication between cars, and perform platoon maneuvers. In this paper, we proposed a driver incentive model which encourages platooning on roads leading to driver safety. As, the leader of platoon have multiple responsibilities than followers, our model rewards more incentives to leader than followers. These incentives will be rewarded as crypto tokens. This digital monetization method for both leaders and followers of a platoon is accomplished by secure transactions using blockchain.

\end{abstract}
\section{Introduction}
In our daily lives, we spend an average of 47 minutes on the road travelling, and the traffic on the roads is becoming increasingly crowded. Congested roadways are linked to longer commute, lower fuel economy, and a higher risk of motor vehicle collisions. We could save a lot of travel time if we could reduce commuting times by a fraction of a second. One solution to the traffic problem is cooperative driving, also known as platooning \cite{bergenhem2012overview}. Vehicles in a platoon communicate using an ad-hoc network or other communication protocols. These communication channels allow platoons to drive closer to one other while maintaining a safe distance. A platoon of vehicles will have a leader who will interact with the platoon followers while managing the platoon and overseeing maneuvers. The platoon leader is in charge of speed, lane changes, braking, and so on, while the follower vehicles are in charge of following the leader vehicle. 

Cooperative driving uses vehicle-to-vehicle and infrastructure-to-vehicle wireless communication system. \cite{piao2008advanced} emphasizes the technology aids in the interchange of data gathered from other cars that is impossible to obtain via on-board sensors. The Advanced Transportation Technology (PATH) project in California \cite{shladover1991automated} first proposed the idea of cars traveling together on the road in 1980. Cooperative driving can improve the driving experience on the road by relieving the driver from some of the driving obligations. Traditional sensor based Adaptive Cruise Control (ACC) isn't enough for cooperative platooning, instead Cooperative Adaptive Cruise Control (CACC) should be considered. CACC broadcasts information such as speed, acceleration, and distance through wireless communication. By allowing CACC, the distance between vehicles can be minimized by following closely, improving both safety and fuel efficiency 
. The focus on cooperative driving or platooning has increased globally 
in recent years because of the potential it holds in road transportation mainly focusing on automated and mixed traffic. Truck platooning \cite{alam2015heavy,liang2016fuel}, and CACC \cite{shladover2015cooperative} were prominent examples of cooperative driving, which focused on minimizing inter-vehicular distance by obeying the "Three Second Rule" safety rule \cite{wolf2008artificial}.

Having a good leader for a platoon is really crucial in forming, maintaining, and improving safety. There are a lot of methods in electing a platoon leader. \cite{singh2018leader} proposed an incentive based strategy using blockchain to elect leader who is the best for the safety of the platoon. The other way is through voting \cite{amoozadeh2015platoon} to elect the platoon leader. Some other methods are may be through scoring and ranking the drivers based on the everyday driving and the driver with best rank can only initiate platoon formation. Our inspiration is drawn from the ranking method. We added an incentive or monetization factor for selecting platoon leader based on rank. When a driver of a vehicle drives everyday there will be a rank assigned to the driver performance and the driver with a best rank can become a platoon leader and other drivers will be followers. To encourage more drivers to be  platoon leaders, a monetization system is required that is fair for all members. This digital monetization is implemented by the usage of a smart contract in blockchain technology that holds users to a higher level of behavior, which is promising in this regard. This smart contract establishes what constitutes acceptable behavior and prevents users from breaking that standard.

The contributions of this paper is organized as follows: Literature Survey on Platooning and Blockchain is discussed in Section 2. The Cooperative Platoon Earnings Methodology is described in Section 3. Algorithms for determining the Earning setup, analysis, scoring model, and monetization heuristic are presented in Section 4. Section 5 discusses the implementation of  Driver Safety Reward (DSR) on Rinkeby Test Network. Conclusions and future work are presented in Section 6.


\section{Literature Review}
The first platooning simulator that was developed was Hestia \cite{halle2005collaborative}. This simulator is used for simulating various scenarios using sensors but its drawback is it do not execute the platooning maneuvers and does not simulate traffic scenarios. In paper \cite{fernandes2010platooning}, the researchers tried to simulate the mixed traffic scenarios using SUMO 
and were unsuccessful. They implemented a car following model based on CACC to simulate inter-vehicle communication
. In \cite{van2006impact} paper, they manually simulated the mixed traffic scenarios and tested the simulator to study the consequences of CACC on traffic. 
 
The PLEXE \cite{segata2014plexe} simulation tool is a platooning extension for SUMO \cite{SUMO2018} which is open-source and is available to the community. This simulation tool has different CACC car-following models to experiment. The wireless communication protocols are available for simulating the formation and platoon management. Mixed traffic scenarios are available to use and implement platooning maneuvers. The authors in paper \cite{kazerooni2015interaction}, developed a simulator to implement platooning maneuvers such as join an existing platoon and merge two platoons. In paper \cite{amoozadeh2015platoon}, the authors developed a state-of-the-art simulator based on VENTOS \cite{amoozadeh2019ventos} which uses SUMO. PERMIT \cite{mena2018permit} is a tool which simulates platooning maneuvers like join, merge, leave, and split which is built on Plexe \cite{segata2014plexe}.

Currently, to our knowledge only few researchers are working on effectively combining the benefits of blockchain technology with the platooning technology to better understand the usefulness. The authors in paper \cite{rowan2017securing} used blockchain as a medium in transportation. They achieved the communication between vehicles in platoon and blockchain public key infrastructure by securely using hardware-based side channels. In \cite{calvo2018secure}, they decreased the blockchain transaction validation time, and verified the vehicle identity. The authors in \cite{hexmoor2018blockchain}, emphasised on using blockchain with platooning to share information securely and rapidly. 

In our preceding work \cite{rachamalla2022dsrbt}, significant features are extracted from a simulated driving dataset and driver is assigned a rank based on the behaviour on road. Based on the rank, the driver is offered crypto tokens and these transaction details along with driver attributes are stored on a blockchain network.


\section{Cooperative Platoon Earnings Methodology}
\begin{figure*}[ht]
\includegraphics[height=5cm, width=1\textwidth]{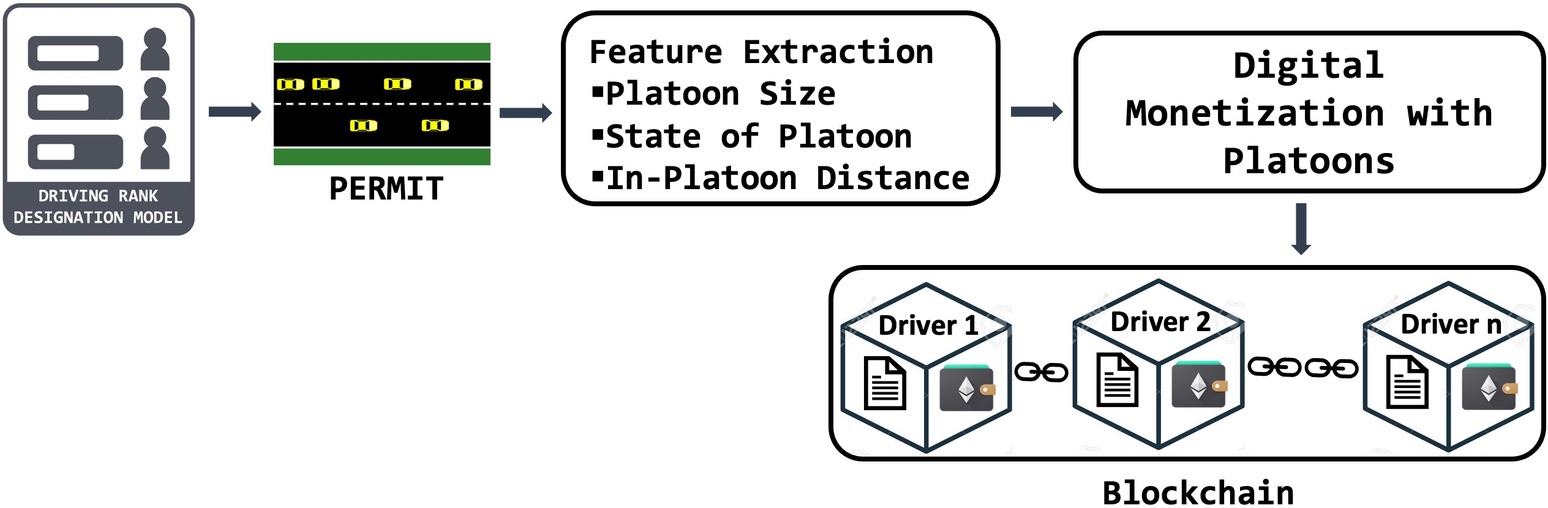}
\caption{Cooperative Platoon Earnings Methodology}
\label{fig:architecture}
\end{figure*}

This section presents the entire overview of the proposed framework as shown in the Fig. \ref{fig:architecture}. This methodology put forward in 5 steps, namely: (1) Driving Rank Designation Model (2) Simulation with Permit (3) Feature Extraction (4) Digital Monetization with Platoons (5) Storing in Blockchain.

\textbf{\textit{Driving Rank Designation Model:}} 
Each Driver is assigned a rank and based on the rank,  Earning rate of the driver is determined by the heuristic as proposed in \cite{rachamalla2022dsrbt}.

\textbf{\textit{Simulation with PERMIT:}}
PERMIT \cite{mena2018permit}, an open source platooning simulator based on SUMO and its platooning extension PLEXE. With PERMIT, Merge, Join, Leave and Split maneuvers can be performed.
\textit{Merge} is a maneuver in which two platoons join to form one platoon.
\textit{Join} refers to joining one vehicle into an existing platoon.
\textit{Leave} maneuver is when a car exists the current platoon.
\textit{Split} refers to dissolving one single platoon into two platoons. By using the PERMIT, we simulated all these maneuvers. This provides the data required for the evaluating the earnings with platoon for a driver.

\begin{figure}[!ht]
\centering
\includegraphics[width=4cm, height=4cm]{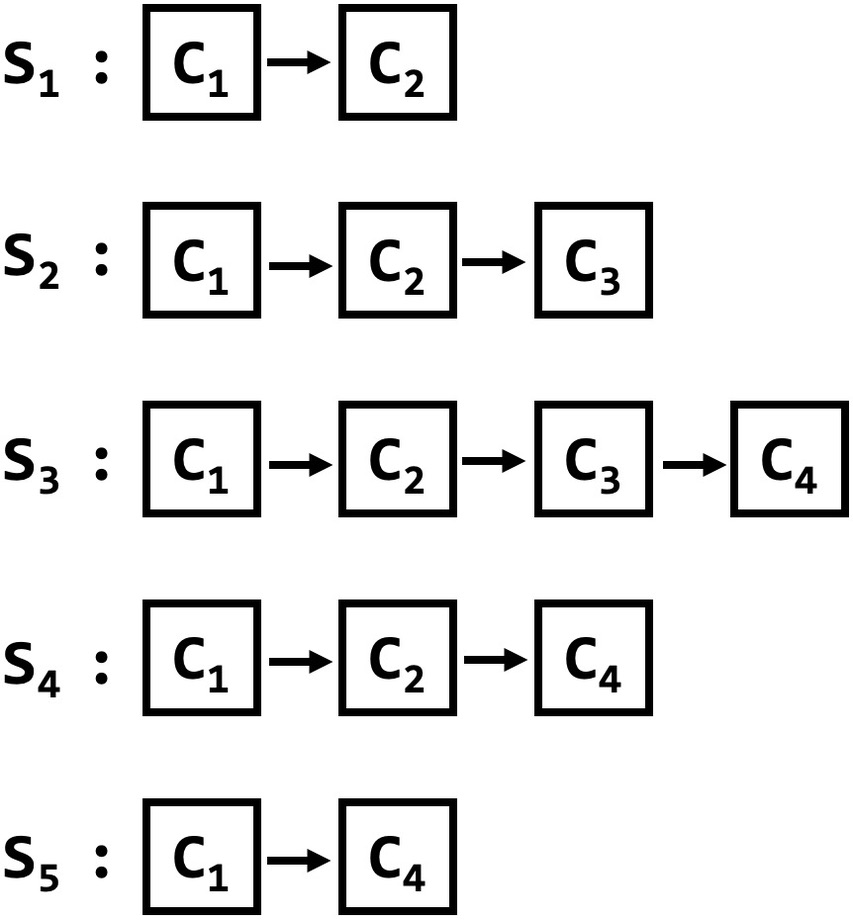}
\caption{States in Platoon}
\label{fig:states}
\end{figure}

\textbf{\textit{Feature Extraction:}}
A platoon state represents two features: (1) number of cars, and (2) distance travelled. For example, in Fig. \ref{fig:states}, a platoon is represented with five different states. First state $S_1$ has two cars. Car $C_3$ has joined the platoon reaching to state $S_2$. Similarly, $S_3$ is achieved. In contrast, $S_4$ is attained by car $C_3$ leaving the platoon. Similarly, $S_5$ is also reached.

\begin{table}[ht]
\centering
    \caption{List of Notations}
    \begin{tabular}{|p{0.15\linewidth} | p{0.75\linewidth}|}
      \hline
\textbf{Symbol} & \textbf{Definition}  \\
\hline
  $er_{d}$ & Earnings of particular day\\
	$er^{out}$ & Earnings outside of the platoon\\
	$er^{in}$ & Earnings inside of the platoon\\  
	$ER_{d-1}$ & Earning Rate of previous day\\
	$d^{out}$ & Out-platoon distance \\
	$n$ & Number of cars in a platoon \\
	$j$ & Number of followers in the platoon in join maneuver \\
	$l$ & Number of cars left the platoon in leave maneuver\\
	$\delta$ & Balancing Factor \\
	$\eta$ & Penalty Factor \\
	$S$ & State-of-platoon \\
	$L_i$ & Length of the platoon at state i \\
	$d^{in}$ & In-platoon distance at state S	\\
	$w$ & Number of platoons a driver travelled\\
    $P_{join}^{er}(L)$ & In-platoon earnings with join maneuver of leader \\
    $P_{join}^{er}(F)$ & In-platoon earnings with join maneuver of follower \\
	$P_{leave}^{er}(L)$ & In-platoon earnings with leave maneuver of leader \\
  \hline
    \end{tabular}
    \label{tab:symbols}
\end{table}

\textbf{\textit{Digital Monetization with Platoons:}} With the features extracted from the previous step, we formalized the earnings ($er_d$) for the driver as summation of the earnings achieved while he drove in platoons ($er^{in}$) and earnings achieved while driving outside the platoon ($er^{out}$)

\begin{equation}
er_d = er^{in} + er^{out}\label{eq1}
\end{equation}

Because different maneuvers can exist inside the platoon, we decomposed the earning offered inside the platoon into addition of earnings during join($P_{join}^{er}$) and leave($P_{leave}^{er}$) maneuver.

\begin{equation}
er^{in} = P_{join}^{er} + P_{leave}^{er}\label{eq2}
\end{equation}

However, the $er^{out}$ is calculated as product of previous day earnings rate ($ER_{d-1}$) which determined by the Driving Rank Designation Model and distance travelled by the driver outside the platoon($d^{out}$).

\begin{equation}
er^{out} = ER_{d-1} * d^{out}\label{eq3}
\end{equation}

As mentioned earlier that a platoon leader will have a little favor by this model due to his responsibilities, Join and Leave Maneuvers are calculated using two different equation one representing Leader and other Follower in the platoon.

\textit{Join Maneuver}
For every platoon, driver joined on a particular day, and for all the states in each platoon, calculate the product of the average of the States of the platoon($S_i$) and the sum of the previous earning rate of the driver($ER_{d-1}$) and $n\delta$. The State($S_i$) is defined as the product of the Length of Platoon($L_i$) at state $i$ and distance travelled inside the platoon($d_i^{in}$).
Here the term $j\delta$ is the additional incentive for the leader of the platoon. It represents the summation of the balancing factor over the number cars joined in the platoon. The balancing factor $\delta$ is used to control the amount of incentive does the drivers will provide during the platoon. We assigned it as 0.01.
\begin{equation} 
P_{join}^{er}(L) = {\sum\limits_{p=1}^w}{\sum\limits_{i=1}^n} S_i * (ER_{d-1} + j\delta) \label{eq4}
\end{equation}

The difference between Leader and Follower is there will be no additional incentives for follower. Instead, only the balancing factor is added to the previous day earning rate($ER_{d-1}$). Platoon Follower doesn't require length of the platoon. So, it just uses the distance travelled in each state.
\begin{equation} 
P_{join}^{er}(F) = {\sum\limits_{p=1}^w}{\sum\limits_{i=1}^n} d_i * (ER_{d-1} + \delta) \label{eq5}
\end{equation}

\textit{Leave Maneuver}
During the Leave Maneuver, for the leader, instead of the number of cars joined($j$), number of cars left($l$) is considered. Additionally, there will be a penalty if a car leaves the platoon before travelling $\eta$ miles. In other words, earnings for the followers will start only after travelling $\eta$ miles. The overhead incurred by changing the structure of the platoon while on the move is the main reason for penalty. Here we considered $\eta=10$.
\begin{equation} 
P_{leave}^{er}(L) = {\sum\limits_{p=1}^w}[{\sum\limits_{i=1}^n} S_i * (ER_{d-1} - l\delta) + penalty_w] \label{eq6}
\end{equation}
where,
\begin{equation} 
S_{i} = L_i * d_i^{in} \label{eq7}
\end{equation}

$$
penalty=\begin{cases}
			(d^{in}-\eta)*\delta, & \text{if $d^{in}<\eta$}\\
            0, & \text{otherwise}
		 \end{cases}
$$

All the notations used in the model as summarized in the table \ref{tab:symbols}.

\textbf{\textit{Storing in Blockchain:}}
For secure transaction of the crypto tokens, the extracted data from the Feature Extraction step is stored in a blockchain technology. Features stored are Driver ID, current earnings, Rank designated for the driver, over speed limit count, distance travelled, number of sharp accelerations, number of sharp decelerations, number of platoons he joined, platoon leader activity, earning date.

To ensure the driver safety on road we propose following rules: 
\begin{itemize}
  \item Drivers with rank less than four cannot act as platoon leaders.
  \item Drivers with rank greater than three are encouraged to be platoon leaders
\end{itemize}

\section{Algorithmic Approach}

In this section, we present the multiple algorithms for calculating the earnings of a day inside a platoon for a driver. In the below pseudo code snippet, total earnings for a driver in platoon is presented.

\begin{minipage}{\hsize}
\lstset{frame=single, framexleftmargin=-1pt,framexrightmargin=-17pt,framesep=8pt}
\begin{lstlisting}[mathescape]
$J^{er} := Join Earnings$
$L^{er} := Leave Earnings$
$Earnings = J^{er} + L^{er}$
\end{lstlisting}
\end{minipage}

The following snippet calculate earnings for a merge or join maneuver for leader or follower.

\begin{minipage}{\hsize}%
\lstset{frame=single, framexleftmargin=-1pt,framexrightmargin=-17pt,framesep=8pt}
\begin{lstlisting}
Join Earnings
if number_cars_joining>1:
    calculate merge_earnings
else
    calculate earnings("leader") 
    or
    calculate earnings("follower")
\end{lstlisting}
\end{minipage}

Similar to above snippet of algorithm, leave maneuver for leader or follower are presented.

\begin{minipage}{\hsize}%
\lstset{frame=single, framexleftmargin=-1pt,framexrightmargin=-17pt,framesep=8pt}
\begin{lstlisting}
Leave Earnings
if number_cars_leaving>1:
    calculate split_earnings
else
    calculate earnings("leader")
    or
    calculate earnings("follower")
\end{lstlisting}
\end{minipage}

merge\_earnings and split\_earnings refers to earnings that can be awarded in the merge and split maneuvers respectively for a driver.  
\begin{algorithm}
\caption{merge\_earnings}\label{alg2}
\begin{algorithmic}[1]
\If{$platoon\_1$ \space  $leader$}
    \State $earnings("leader")$
\EndIf
\If{$platoon\_2$ \space $leader$}
    \State $earnings("follower")$
\EndIf
\If{$platoon $ \space $follower$}
    \State $earnings("follower")$
\EndIf
\end{algorithmic}
\end{algorithm}

\begin{algorithm}
\caption{split\_earnings}\label{alg3}
\begin{algorithmic}[1]
\If{$platoon\_1$\space $leader$}
    \State $penalized\_earnings("leader")$
\EndIf
\If{$platoon\_2$ \space $leader$}
    \State $penalized\_earnings("leader")$
\EndIf
\If{$platoon$ \space $follower$}
    \State $penalized\_earnings("follower")$
\EndIf
\end{algorithmic}
\end{algorithm}

Depending on the driver\_type whether its leader or follower of the platoon, \emph{earnings} algorithm is used for a platoon maneuver especially for join. Similarly \emph{penalized\_earnings} is also defined which is used in the case of the leave maneuver.
\begin{algorithm}
\caption{earnings(driver\_type)}\label{alg1}
\begin{algorithmic}[1]
\Require $P^{er}_{join} = 0, \delta=0.01$
\For w in $num\_of\_platoons$:
\If{$driver = {Platoon\_Leader}$}
    \State $P_{er} = ER_{d-1} + (joined\_cars * \delta)$
    \For s in $num\_of\_states$:
        \State $S_P=S_P+(L_s*d_s)$
    \EndFor
\Else
    \State $P_{er} = ER_{d-1} + \delta$
    \For s in $number\_of\_states$:
        \State $S_P=S_P+(d_s)$
    \EndFor
\EndIf
\State $P_{join}^{er} = P_{join}^{er} + S_P + P_{er}$
\EndFor
\end{algorithmic}
\end{algorithm}

\begin{algorithm}
\caption{penalized\_earnings(driver\_type)}\label{alg4}
\begin{algorithmic}[1]
\Require $P_{leave}^{er} = 0, \delta=0.01$
\For w in $number\_of\_platoons$:
\State calculate $penalty$
\If{$driver = {Platoon\_Leader}$}
    \State $P_{er} = ER_{d-1} - (left\_cars * \delta)$
    \For s in $num\_of\_states$:
        \State $S_P=S_P+(L_s*d_s)$
    \EndFor
\Else
    \State $P_{er} = ER_{d-1} + \delta$
    \For s in $num\_of\_states$:
        \State $S_P=S_P+(d_s)$
    \EndFor
\EndIf
\State $P_{leave}^{er} = P_{leave}^{er} + S_P + P_{er}+penalty$
\EndFor
\end{algorithmic}
\end{algorithm}

\section{Driver Safety Reward in Platoons}\label{sec5}
We implemented a simple six car platoon with Join and Leave Maneuvers in PERMIT. With the data from the PERMIT and the above formulation we calculated the earnings for leader and a follower in both the scenarios. For the leader, the earnings resulted in 26.29 tokens and in the case of follower its 19.34 tokens.
                                                                                            
A Rinkeby test network is used as our Ethereum network for implementation \cite{rachamalla2022dsrbt}. On the rinkeby etherscan network, two smart contracts were deployed, one for tokenization and the other for storing driver records. Tokenization contract will authorize the driver-record contract with driver DSR test tokens and transfer a small number of DSR test tokens to the driver-record contract at first.The driver-record will now be able to credit DSR test tokens to allocated driver's wallets based on their rating. The tokenization contract, for example, created 10000000 ($10^7$) DSR test tokens and authorized the data-record to spend them. The contract sends 10000 DSR test tokens to data-record, where they will be assigned to the drivers.

The tokenization contract, for example, created 10000000 ($10^7$) DSR test tokens and authorized the data-record to spend them. The contract sends 10000 DSR test tokens to data-record, where they will be assigned to the drivers. In the Admin wallet, the MetaMask represents 9990000 (0.999*$10^7$) DSR test tokens. 

\section{Conclusion and Future Work}
In this paper, we proposed a model to incentivize the driver behaviour in the case of platooning. For this, we articulated a system to calculate the earnings awarded for a driver in a platoon. This system considers the different maneuvers in platoon namely Join, Leave, Merge, and Split. We used PERMIT, a platoon simulating system to mimic the aforementioned maneuvers. A Rinkeby test network is used to award the test tokens to driver based on their earnings. This paper assumes for the platooning in cars, but can be extended to different homogeneous vehicles and even heterogeneous vehicles.
%
%
%
\bibliographystyle{unsrt} 
\bibliography{Driver_Safety_Reward_with_Cooperative_Platooning_using_Blockchain.bib}

\end{document}